\documentstyle[preprint,amsfonts,aps]{revtex}
\newcommand{\vR}{\bbox{R}}              
\newcommand{\vrr}{\bbox{r}}             
\newcommand{\va}{\bbox{\alpha}}         
\newcommand{\vb}{\bbox{\beta}}          
\newcommand{\vd}{\bbox{\delta}}         
\newcommand{\vk}{\bbox{k}}              
\newcommand{\vq}{\bbox{q}}              
\newcommand{\ra}{\rightarrow}           
\newcommand{\cf}[1]{\langle #1 \rangle} 
\newcommand{\onehalf}{\mbox{$\frac{1}{2}$}}      
\newcommand{\threehalf}{\mbox{$\frac{3}{2}$}}   
\newcommand{\onefifth}{\mbox{$\frac{1}{5}$}}      
\newcommand{\onetwenth}{\mbox{$\frac{1}{20}$}}    
\newcommand{\Sp}{\ \ }                  
\newcommand{\cS}{{\cal S}}              
\newcommand{\phdagger}{\mathop{\phantom{\dagger}}} 
\newcommand{\chit}{\chi^{ }_{\perp}}               
\newcommand{\bop}[1]{b^{\phdagger}_{#1}}           
\newcommand{\bdop}[1]{b^{\dagger}_{#1}}            
\newcommand{\aop}[1]{a^{\phdagger}_{#1}}           
\newcommand{\apop}[1]{a'^{\phdagger}_{#1}}         
\newcommand{\cop}[1]{c^{\phdagger}_{#1}}           
\newcommand{\cdop}[1]{c^{\dagger}_{#1}}            
\newcommand{\cpop}[1]{c'^{\phdagger}_{#1}}         
\newcommand{\cpdop}[1]{c'^{\dagger}_{#1}}          
\newcommand{\Aop}[1]{{\cal A}^{\phdagger}_{#1}}    
\newcommand{\Bop}[1]{{\cal B}^{\phdagger}_{#1}}    
\newcommand{\Xop}[1]{{\cal X}^{\phdagger}_{#1}}    
\newcommand{\Xdop}[1]{{\cal X}^{\dagger}_{#1}}     
\newcommand{\bsigma}{\mbox{\boldmath $\sigma$}}  
\newcommand{\Neel}{N\'{e}el}                   

\newcommand{\comment}[1]{}

\begin{document}
\draft
\preprint{\vbox{\hbox{\large  \, \ \ \ \ \ \ \ \ G\"oteborg ITP 93-36}
                \hbox{\large cond-mat/9310063}
                \hbox{\large Submitted to Phys. Rev. B}}}

\title{\Large\bf Frustrated Honeycomb Heisenberg Antiferromagnet:
                 A Schwinger Boson Approach}

\author{\large Ann Mattsson, Per Fr\"ojdh}

\address{\normalsize\em Institute of Theoretical Physics\\
Chalmers University of Technology and\\
University of G\"oteborg\\
S-412 96 G\"oteborg, Sweden}

\author{\large Torbj\"orn Einarsson}

\address{\normalsize\em Laboratoire de Physique des Solides\\
Universit\'e de Paris-Sud\\
F-914 05 Orsay, France}

\maketitle
\begin{abstract}
We analyze the frustrated Heisenberg antiferromagnet defined on a honeycomb
lattice using a Schwinger-boson mean-field theory.
The spin-wave velocity and the susceptibility are presented as
functions of the strength of the frustrating interaction for spin
$S=\onehalf$, and the dynamic structure factor is calculated for various
temperatures and frustrations. For large $S$, we find an increased
\Neel\ stability with respect to the classical case.
\end{abstract}

\vspace{4mm}


\narrowtext
\section*{Introduction}
The two-dimensional Heisenberg antiferromagnet has received much
attention during the last years. One of the reasons for this new interest
stems from the discovery of high-$T_c$ superconductors, which
has encouraged
comprehensive investigations of the model defined on a square
lattice. In particular, there is now convincing evidence
that it has a N\'{e}el ordered ground state for nearest-neighbor
interactions even for spin $S=\onehalf$ \cite{Manousakis}.
For the honeycomb lattice, however, the situation is less clear. The
special features of this lattice make it more intricate and it has so
far been paid less attention.

The honeycomb lattice is bipartite and has a \Neel\ state as its classical
ground state. However, since it has the smallest possible two-dimensional
coordination number ($z=3$), quantum fluctuations are expected to be
larger than for a square lattice.
Hence, it can be used for analyzing whether quantum fluctuations may destroy
antiferromagnetic long-range order (AF LRO) in higher dimensions than one.
There are strong indications though that this is not the case,
even for $S=\onehalf$.
Extrapolations of finite-lattice diagonalizations \cite{numerical}
and of Monte Carlo simulations \cite{MC}, as well as first-order corrections
to spin-wave theory \cite{SWT} and series expansions around the Ising
limit \cite{series} all give a finite, albeit much reduced, N\'{e}el-order
parameter. The AF LRO at $T=0$ is also found in our approach.

Another property of the honeycomb lattice is that it is not a Bravais
lattice, i.e., translational invariance of the full lattice is broken for
any type of state. Hence, for a transition from magnetic disorder at finite
temperatures to N\'{e}el order at $T=0$, the spatial symmetry is not
reduced as for the square lattice.
Our treatment here should therefore be possible to extend to a Kagom\'{e}
lattice, which is neither a Bravais- nor a bipartite lattice.
The non-Bravais character has also led to a more
exotic speculation, namely that the model possesses a Haldane gap if
it fails to be N\'{e}el-ordered \cite{H_gap}. However, as just mentioned,
several results are not in favor of this possibility.

After our corroboration of the stability of the \Neel\ state on the
honeycomb lattice at $T=0$, it is of course interesting to consider
the behavior under frustrating interactions. How will the
thermodynamic parameters change, what will happen to the dynamic
structure factor, and when will the \Neel\ state finally break down?

In this paper we calculate several thermodynamic parameters and
investigate effects of frustration of the spin-$\onehalf$ case.
For this purpose we
employ the Schwinger-boson mean-field theory (SBMFT), which has proved
successful in incorporating quantum fluctuations \cite{AA} and which two
of us have used for an investigation of frustration in the square case
\cite{ToPeHe}.
In particular, we report results for the spin-wave velocity and
transverse susceptibility, with and without frustration, at zero
temperature. We also calculate the dependence of frustration for the
mass of spin-wave excitations at low temperature.

While the weak-frustration limit is well-defined and well-behaved
both in spin-wave theory and SBMFT, the case of strong frustration
close to the classical critical frustration is more complicated.
For the square lattice, for example, the location of the
\Neel\ boundary has been an issue of considerable interest and it is
not settled yet. In the present paper we present results for the honeycomb
\Neel\ boundary using SBMFT, linear spin-wave theory (LSWT), and a mapping
to the nonlinear $\sigma$ model with SBMFT parameters calculated here for
$S=\onehalf$. The qualitative differences between the various
methods are the same as for the square lattice.

Finally, we present calculations of the dynamic structure factor, which
allows for a detailed study of the nature of the spin-wave excitations.
In particular, we probe the excitations around the antiferromagnetic
Bragg peak and their dependence on frustration and temperature.

\section*{The model}
The honeycomb lattice can be viewed as composed of two interlacing
triangular sublattices. Each site has its three nearest neighbors on the
other sublattice and its six second-nearest neighbors on its own sublattice,
whereas the third-nearest neighbors are again on the opposite sublattice.
With positive exchange couplings, the nearest-neighbor interactions classically
lead to a N\'eel-ordered ground state, which is frustrated by the
interactions within the sublattices.
In our study of the frustrated quantum Heisenberg antiferromagnet, we
consider the $J_1$-$J_2$ model
\begin{equation}
{\cal H} =
J_{1}\sum_{\vrr,\va}{\bf S}_{\vrr} \cdot
  {\bf S}_{\vrr+\va} +
J_{2}\sum_{\vR , \vb}{\bf S}_{\vR} \cdot
    {\bf S}_{\vR+\vb} \Sp, \label{neighbor}
\end{equation}
with $J_{1}, J_{2} \geq 0$. The first sum runs over all sites $\vrr$ of one
sublattice and over the three nearest-neighbor vectors $\va$.
The next sum, on the other hand, runs over all lattice sites $\vR$ and
three of the six second-nearest-neighbor vectors $\vb$. We have chosen
\begin{mathletters}
\label{alpha}
\begin{eqnarray}
\va_1 & = & \frac{\sqrt{3}}{2} {\bf e}_x + \frac{1}{2} {\bf e}_y \Sp, \\
\va_2 & = & - \frac{\sqrt{3}}{2} {\bf e}_x + \frac{1}{2} {\bf e}_y \Sp, \\
\va_3 & = & - {\bf e}_y\Sp,
\end{eqnarray}
\end{mathletters}
and
\begin{mathletters}
\label{beta}
\begin{eqnarray}
\vb_1 & = & \frac{\sqrt{3}}{2} {\bf e}_{x} - \frac{3}{2} {\bf e}_{y} \Sp, \\
\vb_2 & = & \frac{\sqrt{3}}{2} {\bf e}_{x} + \frac{3}{2} {\bf e}_{y} \Sp, \\
\vb_3 & = & -\sqrt{3}{\bf e}_{x}\Sp,
\end{eqnarray}
\end{mathletters}
and units where the (nearest-neighbor) lattice constant equals 1.
These vectors are depicted in Fig.~\ref{fig:honey}.

The Schwinger-boson approach for the honeycomb lattice is similar to
the one on the square lattice. However, as the former is not a Bravais
lattice, complex phases enter geometric factors and it is necessary to
treat the sublattices separately.
In the following, we give the main formulas for
SBMFT on a honeycomb lattice, which are
to be compared with the corresponding formulas for the square case
(Eqs. (4)--(12) in Ref.~\cite{ToPeHe}).
\comment{More about the differences? Ann?}
The spin operators ${\bf S}_{\vR}$ at each lattice site are replaced by
two species of Schwinger bosons $\bdop{\mu \vR} \ (\mu = 1,2)$ via
\begin{equation}
{\bf S}_{\vR} = {1 \over 2} \bdop{\mu \vR}
{\bsigma}^{\phdagger}_{\mu \nu} \bop{\nu \vR} \ \Sp, \label{julianS}
\end{equation}
with the local constraints
$\bdop{\mu {\vR} } \bop{\mu {\vR} } = 2S.$
Here $\bsigma = ({\sigma}^x, {\sigma}^y, {\sigma}^z)$ is the vector of
Pauli matrices,
and summation over repeated (Greek) indices is implied.
The spins on one sublattice are rotated by $\pi$ around the $y$-axis,
which leads us to the following Hamiltonian:
\begin{equation}
{\cal H} =
  - J_1 \sum_{\vrr, \va} {\cal W}^{\cal A}_{\vrr,\va}
  + J_2 \sum_{\vR, \vb} {\cal W}^{\cal B}_{\vR,\vb}
 + \sum_{\vR } \lambda_{\vR} [\bdop{\mu  \vR} \bop{\mu \vR} -2S] \Sp,
\label{bondH}
\end{equation}
where we have included the local constraints with Lagrange
multipliers $\lambda_{\vR}$ at each site. The expression for the summands is
\begin{equation}
{\cal W}^{\cal X}_{\vR,\vd} \Sp =
   \Sp \onehalf :\Xdop{\vR, \vd} \Xop{\vR, \vd}: - S^2 \Sp,
\label{Wdef}
\end{equation}
with $\Xop{\vR, \vd}$ any of the two link operators
\begin{mathletters}
\begin{eqnarray}
\Aop{\vR, \vd} & \equiv & \bop{1 \vR} \bop{1 \vR+ \vd} +
\bop{2 \vR} \bop{2 \vR + \vd} \ , \label{bondOPa} \\
 \Bop{\vR, \vd} & \equiv & \bdop{1 \vR} \bop{1 \vR + \vd} +
\bdop{2 \vR} \bop{2 \vR + \vd} \ . \label{bondOPb}
\end{eqnarray}
\end{mathletters}
The mean-field theory is finally generated by the Hartree-Fock decoupling
\begin{equation}
:\!\Xdop{\vR, \vd} \Xop{\vR, \vd}\!:  \,\rightarrow
\Xdop{\vR, \vd} \cf{ \Xop{\vR, \vd} } +
\cf{ \Xdop{\vR, \vd} } \Xop{\vR, \vd} -
\cf{ \Xdop{\vR, \vd} } \cf{ \Xop{\vR, \vd} }\Sp,\label{Hartree}
\end{equation}
where the link fields
$Q_1 \equiv \cf{\Aop{\vrr , \va}}$ and
$Q_2 \equiv \cf{\Bop{\vR , \vb}}$
are taken to be uniform and real.
In our mean-field treatment, we replace the local
Lagrange multipliers ${\lambda}_{\vR}$ by a single parameter $\lambda $.

After Fourier-transforming the Schwinger bosons independently on each
sublattice,
\begin{mathletters}
\begin{eqnarray}
\bop{\mu {\vR}} & = & \frac{1}{\sqrt{N/2}} \sum_{\vk}
e^{-i \vk \cdot \vR} \aop{\mu \vk} \Sp , \Sp
\label{Ftransfa} \\
\bop{\mu {\vR'}} & = & \frac{1}{\sqrt{N/2}} \sum_{\vk}
e^{i \vk \cdot \vR'} \apop{\mu \vk} \Sp , \Sp
\label{Ftransfb}
\end{eqnarray}
\end{mathletters}
with $\vR$ and $\vR'$ on different sublattices,
we use a (complex) Bogoliubov transformation
\begin{mathletters}
\begin{eqnarray}
\aop{\mu \vk} & = & e^{-i \varphi_{\vk}} \cosh{{\vartheta}_{\vk}} \cop{\mu \vk}
+ \sinh{{\vartheta}_{\vk}} \cpdop{\mu \vk} \Sp , \label{Btransfa} \\
\apop{\mu \vk} & = & e^{-i \varphi_{\vk}} \cosh{{\vartheta}_{\vk}}
\cpop{\mu \vk}
+ \sinh{{\vartheta}_{\vk}} \cdop{\mu \vk} \Sp , \label{Btransfb}
\end{eqnarray}
\end{mathletters}
to diagonalize the resulting Hamiltonian by choosing $\varphi_{\vk}$ and
${\vartheta}_{\vk}$ properly. This yields free bosons with dispersion relation
\begin{equation}
\omega_{\vk } = \sqrt{(\lambda + 3 J_2 Q_2 \gamma_{2,\vk} )^2 -
|\threehalf J_1 Q_1 \gamma_{1,\vk} |^2}  \Sp,
\label{dispersion}
\end{equation}
where the geometrical factors are given by
\begin{mathletters}
\begin{eqnarray}
\gamma_{1, \vk} & = & \frac{1}{3}
\sum_{\va} e^{i \vk \cdot \va} \Sp,\\
\gamma_{2, \vk} & = & \frac{1}{3}
\sum_{\vb} \cos ( \vk \cdot \vb ) \Sp.
\end{eqnarray}
\end{mathletters}

The mean-field equations are the minimization equations
for the free energy:
\begin{mathletters}
\label{INTEQ} 
\begin{eqnarray}
\int \frac{d^2 k}{A} \cosh (2 {\vartheta}_{\vk})
  (n_{\vk }+ \onehalf) - (S+\onehalf)  &=& 0 \Sp,\label{INTEQa}\\
\int \frac{d^2 k}{A} \sinh(2 {\vartheta}_{\vk} )
|\gamma_{1, \vk}|(n_{\vk } + \onehalf) - \onehalf
 Q_1  &= &  0 \Sp ,   \label{INTEQb}\\
\int \frac{d^2 k}{A} \cosh(2 {\vartheta}_{\vk} )
\gamma_{2, \vk}(n_{\vk } + \onehalf) - \onehalf
 Q_2  &= &  0 \Sp ,   \label{INTEQc}
\end{eqnarray}
\end{mathletters}
with $A=8 \pi^{2}/(3 \sqrt{3})$ the area of the reciprocal unit cell of
a sublattice, $n_{\vk }= [\exp(\beta\omega_{\vk})-1]^{-1}$ being the
Bose occupation number and
\begin{equation}
{\rm tanh} (2{\vartheta}_{\vk}) =
\frac{\threehalf J_1 Q_1 | \gamma_{1,\vk}|}
{\lambda + 3J_2 Q_2 \gamma_{2,\vk} } \Sp.
\label{angle}
\end{equation}
Solving the mean-field equations yields values for
$Q_1$, $Q_2$, and $\lambda$, and also the opportunity to determine
thermodynamic quantities at finite temperatures.

In two dimensions there can be LRO only at zero temperature. This
means that exactly at $T=0$ there is an abrupt phase transition,
and we must be careful when reducing $T$ to zero. As the AF LRO
corresponds to a condensation of the Schwinger bosons, we obtain the
mean-field equations at $T=0$ from Eqs.~(\ref{INTEQ}) by the
replacement
\begin{equation} \label{eq:T=0-replacement}
\frac{n_{\vk}}{\omega_{\vk}} \ra S^{*}\delta^{(2)}(\vk) \Sp,
\end{equation}
where $S^{*}$ is a new unknown quantity measuring the Bose condensate.
On the other hand, we can now determine $\lambda$ from $Q_1$ and $Q_2$, as
there is no gap in the spin-wave spectrum ($\omega_{\vk=0}=0$
in (\ref{dispersion})) at $T=0$.

\section*{Thermodynamic parameters}
For small $k$ the dispersion relation (\ref{dispersion}) takes a
relativistic form:
 ${\omega}_{\vk } = c\sqrt{(mc)^2 + |{\vk}|^2}$, with the spin-wave
velocity $c$ being
\begin{equation}
c =\frac{3}{2} \sqrt{\onehalf J_1^2 Q_1^2 - 2J_2 Q_2 (\lambda + 3J_2 Q_2 )}
\Sp , \label{c}
\end{equation}
and a mass $m$ in the energy gap $\Delta=mc^2$ of the
spin-wave excitations:
\begin{equation}
\Delta=mc^2 = \sqrt{(\lambda + 3J_2 Q_2 )^2 - (\threehalf J_1 Q_1)^2} \Sp.
\label{m}
\end{equation}

We have calculated the spin-wave velocity $c$ at $T=0$ with the
zero-temperature formalism and by extrapolation from finite $T$.
The results of the two methods are the same within the accuracy of the
extrapolation, and they are shown
in Fig.~\ref{fig:c} together with the result for LSWT.
Without frustration one can compare our result with the first-order SWT
(FSWT) obtained from $1/S$-corrections to LSWT. In fact, they give
exactly the same result
\begin{equation}
 c=1.2098417*c_{\rm LSWT}=1.2832309 J_1\Sp.
\end{equation}
\comment{This is probably not true for frustration since $c_{FSWT}$
probably vanishes
at the classical frustration.
Would be nice to have something more
substantial here. Is there any formula for $c$ for FSWT with
frustration in the square case? Most discussion in Ferrer for the
square case. Can one learn something from him.}

Next we consider the gap, or the mass $m$.
As in the square case \cite{Manousakis}, without
frustration it behaves like $m \propto \exp(-\mbox{const}/T)$ as $T \ra 0$.
\comment{Is this true?}
For fixed temperatures the mass (=the gap) varies along with frustration.
As can be seen in Fig.~\ref{fig:m}, it increases nearly exponentially with
frustration, in analogy with the square case.

Our last thermodynamic parameter is the
uniform transverse susceptibility $\chi_\perp$. To calculate this, we
first obtain the rotational average $\bar \chi$
at finite $T$ from the static structure factor
$\cS_{xx}({\vq=0}, t=0)$ by
\begin{equation}
{\bar {\chi}} = \frac{1}{k_B T}
\cS_{xx}({\vq }= 0, t=0) \Sp, \label{susc}
\end{equation}
where
\begin{equation}
\cS_{xx} ({\vq } = 0, t=0 ) = \frac{2}{3}
\int\frac{d^2k}{A}\onehalf n_{\vk}(n_{\vk}+1)\Sp. \label{Sfactor}
\end{equation}
Here, the factor $\frac{2}{3}$ has been inserted on the right-hand side of
(\ref{Sfactor}) to make up for the
mean-field treatment of the local constraints \cite{AA}.
Second, we extrapolate $\bar\chi$ down to $T=0$, and use that at $T=0$ the
longitudinal susceptibility $\chi_\parallel \equiv 0$, so that
$\chit = \frac{3}{2}{\bar {\chi}}$.
\comment{I am not sure about this 3/2 any
longer. This argument sounds wrong to me. By the way, note that the
SBMFT result is 1.460*$\chi_{FSWT}$.}
 The result is shown in Fig.~\ref{fig:chi},
where one can also see that the LSWT is independent of frustration.
The extrapolation scheme reduces the precision of the $T=0$ values.
For a test, we have evaluated the exact SBMFT result at $T=0$ with no
frustration by using a formula obtained for the square lattice
\cite{Takahashi}, with integrals calculated for the honeycomb
lattice \cite{SWT}. It reads
\begin{equation}
 \chi_\perp = 0.3997332/6J_1 \Sp ,
\end{equation}
to be compared with our extrapolation
$\chi_\perp = (0.41 \pm 0.01) /6J_1$. Two other results to
compare with are the FSWT result \cite{SWT} $\chi_\perp = 0.274/6J_1$ and
the series result \cite{series} $\chi_\perp = 0.454/6J_1$, which shows
that the SBMFT gives a result close to the one obtained by a series
expansion \cite{series}.
\comment{FSWT:$\chi_\perp= 0.0456287/J_1$,
series: $\chi_\perp= 0.0756(10)/J_1$}

\section*{Stability of the N\'{E}El state}
The classical \Neel\ order vanishes as the frustration
$\alpha \equiv J_2/J_1$ reaches the critical value $\alpha_{\rm cl}=1/6$,
above which each triangular sublattice becomes ordered and totally
decoupled from the other sublattice.
\comment{Exactly at the point $\alpha=1/6$, one may have more
degenerate states. According to the literature
Refs.~{Chubukov,Ferrer}  the case $J_3=0$ is very special in the
square case. Do we have the same thing here? May this be more
investigated using FSWT a la Ferrer?}

The LSWT correction to the classical \Neel\ boundary for the square
lattice was obtained by Chandra and Dou\c{c}ot \cite{ChanDouc} by
finding the frustration at which the sublattice magnetization vanishes.
A repetition of their calculation reduces the \Neel\ stability also in
the honeycomb case, as shown in Fig.~\ref{fig:critS}.
The square-case reduction was soon challenged by modified
spin-wave theory calculations (a theory very similar to SBMFT)
 in Refs.~\cite{BADguys}
and by SBMFT results \cite{MilaPoilBrud}, which instead pointed to an
increase of the \Neel\ region. The discrepancy between SWT and
SBMFT-like theories has later been argued to exist only in LSWT,
and should vanish in FSWT through the cancellation of two logarithmic
divergences \cite{Chubukov,BrudMila,Ferrer}.
Indeed, for large $S$, FSWT and SBMFT give the same magnetization
\cite{BrudMila}. There seems to be general consensus that for large
$S$ there is a stabilization of \Neel\ order beyond the classical
\Neel\ boundary: quantum fluctuations stabilize a state which is
classically forbidden.

We have also calculated the Neel\ boundary by using our $T=0$
mean-field equations. This was done by determining the frustration at which
the Bose condensate amplitude $S^{*}$, and hence the sublattice magnetization,
vanishes. This curve is also presented in Fig.~\ref{fig:critS}.
As in the square case, it is stabilized beyond the classical region, and
enhances the \Neel\ stability even for $S=\onehalf$.
However, in this latter limit of $S=\onehalf$,
the phase boundary obtained by the large-$S$
theories can be trusted less. The most extensive exact
diagonalizations of the $J_1-J_2$ model with extrapolations to the
thermodynamic limit indeed reduces the \Neel\ stability
\cite{SchulzZiman}.

In another approach, we use a mapping to the nonlinear $\sigma$ model
with quantum corrected thermodynamic parameters for the frustrated
spin-$\onehalf$ antiferromagnet on a square lattice \cite{ToHe,ToPeHe}.
The method considerably reduces the \Neel\ stability in
that case. In Ref.~\cite{ToHe} the formulas for the honeycomb lattice
were also given, but at that time the quantum corrections were not
known. Using the values obtained here, these formulas yield
that already without frustration the bare coupling $\tilde{g}_0$ of the
nonlinear $\sigma$ model exceeds the fixed point value $g_c$,
\begin{equation} \label{eq:nlsm}
\tilde{g}_0 =(8\pi\sqrt{3})^{1/2}
              \frac{c_{\rm LSWT}}{c_{\rm SBMFT}}
              \frac{\chi_{\rm LSWT}}{\chi_{\rm SBMFT}} \approx
               13.64 > g_c = 4\pi \approx 12.57\Sp,
\end{equation}
assuming the hydrodynamic relation $\rho=\chi c^2$. \comment{[using the right
units]} This means that the model is in its massive,
disordered state, and that for $S=\onehalf$ there is no \Neel\ order even
without frustration. This result contradicts the ones obtained previously,
and most likely this method underestimates the \Neel\ stability.
Therefore, the result obtained in Ref.~\cite{ToPeHe} for the critical
frustration on the square lattice is most probably too low as well, although we
believe the tendency of a reduced \Neel\ stability for
$S=\onehalf$ to be valid both on the square and the honeycomb lattice.

To sum up, qualitatively SBMFT and LSWT give the same \Neel\ boundaries
as on the square lattice, while the effective-action method
underestimates the \Neel\ stability for $S=\onehalf$. For $S$ large,
the SBMFT phase boundary probably can be trusted, but much work remains
to be done to determine the disordering phase transition for
the $S=\onehalf$ honeycomb antiferromagnet.

\comment{For the square lattice we have the following very odd result for the
\Neel\ boundary from FSWT (use formulas in CHN). At $J_2=0$ the
result is the same as LSWT, but then the critical spin decreases with
the frustration (it doesn't go to zero) and above $J_2 = 0.1$ there is
no critical spin --- there is always order.
This strange effect has to do with that the SWT
expression for the sublattice magnetization is an alternating series in
$1/S$, and for FSWT the last term is positive and stabilizes the \Neel\
state for large enough $1/S$. At $1/S$ very close to zero the
sublattice in FSWT agree with that in SBMFT. The same things should
apply to the honeycomb lattice. Ferrer instead looks at the spin
stiffness, and claims to be able to see the order-from-disorder
phenomena in FSWT. It would be interesting to redo his calculations
for the honeycomb lattice.}

\section*{The dynamic structure factor}
Finally we examine the spin dynamics by means of the dynamic structure
factor $\cS(\vq,\omega)$. For $T\neq0$ there is no LRO, but local
\Neel\ order still supports spin-waves with not too-long wavelengths.
For finite temperatures, the structure factor is rotationally invariant
and can be  written as
\begin{equation}
\cS_{xx} ({\vq },t) = \sum_{\vR,\vR'} e^{i \vq \cdot (\vR - \vR')}
  \langle S_{\vR}^x(t) S_{\vR'}^x(0) \rangle \Sp,
\end{equation}
where $\vR$ and $\vR'$ run over {\em all\/} lattice sites. Notice that since
$\vR - \vR'$ is not always a multiple of the lattice vectors $\vb$, this
is not a spatial Fourier transform as it is for a Bravais lattice.
The (time) Fourier transform is a sum of two terms
\begin{equation}
\cS_{xx} ({\vq },\omega) =
    \int dt \ e^{i \omega t}{\cal S}_{xx} ({\vq },t) =
  \cS_1({\vq },\omega) + \cS_2({\vq },\omega) \Sp,
\end{equation}
which evaluate to
\begin{mathletters}
\label{struct}
\begin{eqnarray}
\cS_1({\vq },\omega) & = & \frac{1}{2} \int \frac{d^2 k}{A}
\frac{( f_{\vk,\vk+\vq} +1)}{2} n_{\vk}(n_{\vk+\vq} + 1)
\delta({\omega}_{\vk}-{\omega}_{\vk+\vq}+\omega) \Sp, \label{struct1} \\
\cS_2({\vq },\omega) & = & \frac{1}{4} \int \frac{d^2 k}{A}
\frac{( f_{\vk,\vk+\vq} -1)}{2} (n_{\vk}+\Theta (\omega))(n_{\vk+\vq}
 + \Theta (\omega))
\delta({\omega}_{\vk}+{\omega}_{\vk+\vq}-|\omega|) \Sp, \label{struct2}
\end{eqnarray}
\end{mathletters}
where
\begin{equation}
f_{\vk,\vk+\vq}=\cosh (2 {\vartheta}_{\vk}) \cosh (2 {\vartheta}_{\vk+\vq})
- \cos ( \varphi_{\vk} - {\varphi}_{\vk+\vq}) \sinh (2 {\vartheta}_{\vk})
\sinh (2 {\vartheta}_{\vk+\vq})\Sp,
\end{equation}
$\Theta (\omega)$ is the Heavyside step function,
and ${\varphi}_{\vk}$ is the phase of $\gamma_{1,\vk}$. The non-Bravais
character of the honeycomb lattice is reflected by the non-zero value of
${\varphi}_{\vk}$.

The two terms $\cS_1$ and $\cS_2$ correspond to different physical processes
and give rise to two distinct peaks in the structure factor.
In neutron-scattering terms, the first term $\cS_1$
describes a simple scattering process of a neutron against a spin-wave,
while the other term, $\cS_2$, corresponds to a scattering process where
two spin waves are created or annihilated.
(However, this double-peak structure may be difficult to resolve
experimentally at low temperatures. At higher temperatures the $\cS_2$ peak
may be too small and smeared out to be detected.)
\comment{Seems easy to resolve from fig 6.b, since the the gap
moves the S2 peak away.}

Our results are presented in two series of plots in
Figs.~\ref{fig:structvsfrust} and \ref{fig:structvstemp} for
$\vq$ close and parallel to the \Neel\ ordering vector
\begin{equation}
\vq_0=(\frac{2\pi}{\sqrt{3}}, \frac{2\pi}{3})\Sp.
\end{equation}
(Compare with plots for the square case in Ref.~\cite{MilaPoilBrud}.)
All (but one) contain two ``peaks'', where one stems from $\cS_1$ and
the other from $\cS_2$ as indicated in Fig.~\ref{fig:structvsfrust}a.

In Fig.~\ref{fig:structvsfrust}, $T=\onefifth J_1$ (constant)
and we vary frustration (\ref{fig:structvsfrust}a: $J_2/J_1=0$ and
\ref{fig:structvsfrust}b: $J_2/J_1 = 0.1$ $<$
critical frustration).
Near the ordering vector we get ``overdamped'' spin-waves that turn
into a magnetic Bragg peak as $T$ is lowered. Outside this region,
we have the normal spin-wave dispersion
($\cS \propto\delta(\omega - c |\vq|)$).
Note that $\cS_2$ has a small gap in the dispersion. (It is actually a double
gap, as two spin waves are created.) The gap increases along with frustration,
as can be seen when comparing Figs.~\ref{fig:structvsfrust}a and
{}~\ref{fig:structvsfrust}b. (The increase of the gap with frustration
has also been depicted in Fig.~\ref{fig:m}.) Above the gap,
SWT-type dispersion is back to normal.

In Fig.~\ref{fig:structvstemp}, $J_2/J_1 = 0.22$  $>$ the critical
frustration. Being on the disordered side
of the critical frustration, there is no Bragg peak at $T=0$.
This remedies the divergencies in the structure factor and makes it
possible to calculate the structure factor at zero temperature within
our formalism.
In Fig.~\ref{fig:structvstemp}a the results for $T=0$ are shown.
Here, the $\cS_1$-peak has vanished since there are no thermally
excited spin-waves to scatter against. The $\cS_2$ process still gives
a peak due to creation of spin-wave pairs.
At a temperature increase, the $\cS_1$ peak reappears as can be seen
in Fig.~\ref{fig:structvstemp}b.
\section*{Conclusions}
SBMFT works well for calculating thermodynamic properties for
the Heisenberg antiferromagnet on a honeycomb lattice, although this
is not a Bravais lattice. For $S=\onehalf$,
the spin-wave velocity and
the transverse susceptibility both decrease slowly with frustration,
while the excitation gap grows nearly exponentially with increasing
frustration.
For large~$S$ AF LRO is found to be stabilized beyond the
classical boundary. The case of small $S$ calls for further
investigation.  Finally, the dynamic structure factor shows a
double-peak structure with a gap for finite temperatures. In general,
the qualitative behavior is the same as for the square lattice.

\section*{Acknowledgments}
We thank H. Johannesson and A. Chubukov for discussions
and valuable remarks. T.E. acknowledges financial support from the
Swedish Natural Research Council (NFR).


\begin{figure}
\caption{Honeycomb lattice with one of the two triangular sublattices
marked. The vectors $\protect{\va}$ and $\protect{\vb}$ point to the nearest
neighbors and half the number of second-nearest neighbors of a site,
respectively. \label{fig:honey}}
\end{figure}

\begin{figure}
\caption{The $T=0$ spin-wave velocity vs frustration
$J_2/J_1$ for $S=\onehalf$.
The full line shows the SBMFT result and the dashed line the LSWT result.
\label{fig:c}}
\end{figure}

\begin{figure}
\caption{The logarithm of the mass vs frustration for SBMFT at
temperature $T=\onefifth J_1$ and $S=\onehalf$. \label{fig:m}}
\end{figure}

\begin{figure}
\caption{The $T=0$ uniform transverse susceptibility vs
frustration  for $S=\onehalf$.  The full and dashed lines correspond
to SBMFT and LSWT, respectively.\label{fig:chi}}
\end{figure}

\begin{figure}
\caption{The inverse of the critical spin $S_c$ as a function of frustration.
The lines show the LSWT and the SBMFT result. They coincide for
$J_2=0$ and for $J_2/J_1= 1/6$ at the classical point $S=\infty$. Note
the \Neel\ stability beyond the classical value in the SBMFT case.
\label{fig:critS}}
\end{figure}

\begin{figure}
\caption{The dynamic structure factor at a low temperature $T=\onefifth J_1$
close to the ordering vector $\protect{\vq_0}$. In (a) the frustration is
$J_2/J_1 = 0$, and in (b) it is $J_2/J_1=0.1$ $<$ critical frustration.
The peak corresponding to
$\cS_2$ has a gap (starts at finite $\omega$) which increases with
frustration.
\label{fig:structvsfrust}}
\end{figure}

\begin{figure}
\caption{The dynamic structure factor for $J_2/J_1 = 0.22$, which is
a value above the critical frustration. In (a) the result for $T=0$ is shown,
while in (b) we have $T=\onetwenth J_1$.
Note that the $\cS_1$ peak has vanished in (a) due to the lack
of thermally excited spin waves.
\label{fig:structvstemp}}
\end{figure}
\end{document}